\documentclass[twocolumn,showpacs,preprintnumbers,amsmath,amssymb]{revtex4}

\usepackage{graphicx}
\usepackage{dcolumn}
\usepackage{bm}

\begin{document}
  \title{Regularization of fields for self-force problems in curved spacetime: \\foundations and a time-domain application}
  \author{Ian Vega}
  \author{Steven Detweiler }
  \affiliation{Institute for Fundamental Theory, Department of Physics, University of Florida, Gainesville, FL 32611-8440}
  \email{vega@phys.ufl.edu}
  \date{January 15, 2008}

\begin{abstract}

We propose an approach for the calculation of self-forces, energy fluxes
and waveforms arising from moving point charges in curved spacetimes. As
opposed to mode-sum schemes that regularize the self-force derived from
the singular retarded field, this approach regularizes the retarded field
itself. The singular part of the retarded field is first analytically
identified and removed, yielding a finite, differentiable remainder from
which the self-force is easily calculated. This regular remainder solves a
wave equation which enjoys the benefit of having a non-singular source.
Solving this wave equation for the remainder completely avoids the
calculation of the singular retarded field along with the attendant
difficulties associated with numerically modeling a delta function source.
From this differentiable remainder one may compute the self-force, the
energy flux, and also a waveform which reflects the effects of the
self-force.

As a test of principle, we implement this method using a 4th-order (1+1)
code, and calculate the self-force for the simple case of a scalar charge
moving in a circular orbit around a Schwarzschild black hole. We achieve
agreement with frequency-domain results to $\sim 0.1\%$ or better.
\end{abstract}

\maketitle

\newcommand{\eg}{{\it e.g.,\ }}
\renewcommand{\SS}{{\text{S}}}
\newcommand{\R}{{\text{R}}}
\newcommand{\eqn}[1]{{Eq.~(\ref{#1})}}
\newcommand{\beq}{\begin{equation}}
\newcommand{\eeq}{\end{equation}}
\newcommand{\calR}{{\mathcal{R}}}
\newcommand{\calM}{{\mathcal{M}}}

\section{Introduction}

With the advent of gravitational wave astronomy approaching, the
development of accurate and efficient models for gravitational wave
sources has steadily progressed. The ability to predict gravitational wave
amplitudes and waveforms for expected sources will greatly enhance the
usefulness of detectors such as VIRGO, LIGO, and LISA. An interesting
class of relevant sources includes a large $10^2$--$10^{10}\text{M}_\odot$
black hole in a binary with a closely orbiting stellar-mass compact
object.

The orbit and inspiral of a compact object into a substantially more
massive black hole presents a complication for traditional numerical
analysis. A numerical grid must be fine enough to resolve the geometry in
the vicinity of the small object, where the metric appears to be that of
the compact object with tidal distortions from the large hole. But the
grid must be coarse enough to reach the wave-zone of the binary, so
that the waveforms might be carefully monitored. In addition, the
timescale for the effect of radiation reaction is long compared with the
orbital period and the waveform from many orbits will be used in the data
analysis. Together the dramatically different length scales coupled with
the dramatically different time scales present a formidable challenge for
the study of an extreme mass ratio inspiral (EMRI).

Perturbative analysis appears more feasible for the EMRI problem. The
orbiting compact object is modeled as a point mass whose motion, to lowest-order,
approximately follows a geodesic in the background geometry of the large black hole
companion. Thus, the emitted gravitational waves can be calculated reasonably well
using the mature perturbation theory of black holes \cite{ReggeWheeler, Zerilli, Teukolsky73}.
But in this approach it is necessary to go beyond the approximation of geodesic motion
in the background geometry. The effects of radiation reaction on the orbital phase
requires an extension of the usual perturbation analysis to include what is often
called the self-force.

In a so-called frequency-domain approach, one chooses to Fourier decompose
the source and the field and then solves for each Fourier mode of the
field independently. This method works well for a flux calculation if the
spectrum is simple, such as that of a particle in a circular orbit. However
for generic trajectories, including particularly those which reflect the
effects of radiation reaction, the frequency spectrum is complicated
enough to make the frequency-domain analysis numerically expensive.

Further, the field of the particle is singular at the particle's location,
and some regularization procedure is required to calculate self-force
effects. The usual procedure to date is termed \textit{mode-sum
regularization}, as initially developed by Barack and Ori
\cite{BarackO2000, BMNOS02, MinoNS2002}. This regularization prescription
depends crucially upon a decomposition of the \emph{derivatives of the field} into
angular modes, such as spherical harmonics, which are individually finite.
From each finite mode, the part that contributes to the singularity but
not the self-force is identified and removed; the remainders of the field
derivatives for each mode are then summed to determine the finite effect
of the particle's field on its own motion. Generally the sum has convergence
which is only polynomial in the mode number and, thus, requires analysis at
high mode numbers \cite{Lousto00, Burko00, DetweilerMW2003, RiveraMWD2004,
Hikida2004, Hikida2005, HaasP2006, Haas2007, BarackSago2007} to obtain
accurate results.

In this manuscript we introduce a general method for analyzing the field of
a point charge orbiting a black hole and for directly determining the
waveforms and flux integrals as well as the instantaneous self-force acting
back on the charge itself, which includes all of the effects of radiation
reaction in a natural manner.

The strength of this approach lies in the derivation of a wave equation
for a regular field $\psi^R$ which is identical to the retarded field in
the wave zone and whose derivatives at the charge determine the
self-force. We call the determination of $\psi^R$, \textit{field
regularization.} Solving our effective wave equation requires neither
Fourier decomposition in time, nor any angular decomposition for treating
the dramatically different length scales in the EMRI problem, and
circumvents the need for ever calculating the actual singular retarded
field.

The source $S_{\text{eff}}$ of the effective wave equation follows from a
local analysis of the singular part $\psi^S$ of the retarded field,
\eqn{eq:psiS}. Importantly, this effective source is smooth everywhere
except for its limited differentiability at the location of the charge.
Ample freedom in choosing $S_{\text{eff}}$ allows the source to spread out
over a region with a length scale comparable to the size of the black hole
or even to the distance from the charge to the black hole.

We shall describe our approach in terms of a point source with a scalar
charge interacting with its own scalar field while orbiting a large black
hole. The formal extension of these ideas to pure gravity with a small
Schwarzschild black hole perturbing the geometry of a much larger black
hole is completely straightforward at the perturbative level. The details
of this extension to gravity are algebraically complicated but
conceptually simple and will be the focus of a future report.

\section{Organization of this Paper}
\label{sec:organization}

The main objective of this work is to provide a proof of principle for the
process of field regularization described in \S\ref{sec:formalism} as a
time-domain technique for self-force calculation. It verifies that we are
able to achieve results comparable to that obtained with frequency-domain methods
\cite{DetweilerMW2003}, or other time-domain methods relying on the mode-sum
decomposition \cite{ HaasP2006, Haas2007}.

In \S\ref{sec:finitedifference} we describe the details of our numerical
implementation of field regularization applied to a scalar charge in a
circular orbit in Schwarzschild. Tests of the internal consistency of the
numerical implementation are in \S\ref{sec:diagnostics}.

Section \ref{sec:results} displays the results of our self-force analysis
for a scalar charge in  circular orbits at Schwarzschild radius $R=10M$
and $12M$. The time and radial components of the self force are compared
with results from a frequency-domain analysis. We also reconstruct the
entire retarded field and compare this with the retarded field of the
frequency domain analysis.

The discussion in \S\ref{sec:summary} summarizes our results and describes
the strengths of field regularization in comparison with other methods of
self-force calculation and also with methods of current interest for
calculating energy fluxes and waveforms for generic orbits about black
holes.

Appendix \ref{sec:THZ} gives some details of the expansion of the singular
field $\psi^S$ about the point charge and describes how higher order terms in
the expansion increase the overall efficiency of field regularization.

\section{Field regularization}
\label{sec:formalism}
\label{sec:regularization}

For a scalar charge, the general strategy for computing the self-force
first involves solving the minimally-coupled scalar wave equation with a
point charge $q$ source,
\begin{equation}
 \nabla^a\nabla_a\psi^{\text{ret}} = -4\pi q \int_\gamma \delta^{(4)}(x-z(\tau)) d\tau,
 \label{eq:wave}
\end{equation}
for the retarded field $\psi^{\text{ret}}$. Here $\nabla_a$ is the
derivative operator associated with the metric $g_{ab}$ of the background
spacetime and $\gamma$ is the worldline of the charge defined by
$z^a(\tau)$ and parameterized by the proper time $\tau$. The physical
solution of the resulting wave equation will be a retarded field that is
singular at the location of the point charge. As such, a self-force
naively expressed as
\begin{equation}
F_a(\tau)=q\nabla_a\psi^{\text{ret}}(z(\tau))\label{eq:naive}
\end{equation}
will need a regularization prescription to make sense. Early
regularization prescriptions \cite{DeWittB1960, MinoST1997, QuinnW1997}
were based upon a Hadamard expansion of the Green function, and showed
that for a particle moving along a geodesic the self force could be
described in terms of the particle interacting only with the ``tail'' part
of $\psi$, which is finite at the particle itself. Later
\cite{DetweilerW2003} it was realized that the singular part of the field
$\psi^S$ which exerts no force on the particle itself could be identified
as an actual solution to \eqn{eq:wave} in a neighborhood of the particle.
A formal description of $\psi^S$ in terms of parts of the retarded Green's
function \cite{DetweilerW2003} is possible, but generally there is no
exact functional description for $\psi^S$ in a neighborhood of the
particle. Fortunately, an intuitively satisfying description for $\psi^S$
results from a careful expansion about the location of the particle:
\beq
  \psi^S = q/\rho + O(\rho^3/\calR^4) \text{ as } \rho\rightarrow0,
\label{eq:psiS}
\eeq
where $\calR$ is a constant length scale of the background geometry and
$\rho$ is a scalar field which simply satisfies $\rho^2=x^2+y^2+z^2$ in a
very special Minkowskii-like locally inertial coordinate system centered
on the particle, first described by Thorne, Hartle and Zhang
\cite{ThorneHartle85, Zhang86} and applied to self-force problems in
Refs.~\cite{det01, DetweilerMW2003, Detweiler05}.  Not surprisingly
the singular part of the field, which exerts no force on the particle
itself, appears as approximately the Coulomb potential to a local observer
moving with the particle.

Our proposal for solving \eqn{eq:wave}, and determining the self-force
acting back on the particle now appears elementary. First we define
\beq
  \tilde\psi^S \equiv q/\rho
\eeq
as a specific approximation to $\psi^S$. By construction, we know that
$\tilde\psi^S$ is singular at the particle and is $C^\infty$ elsewhere.
Also, within a neighborhood of the worldline of the particle
\begin{eqnarray}
 \nabla^a\nabla_a\tilde\psi^S &=& -4\pi q \int_\gamma \delta^{(4)}(x-z(\tau))\, d\tau + O(\rho/\calR^4),
\nonumber\\ &&  \text{ as }\rho\rightarrow0  .
 \label{eq:waveS}
\end{eqnarray}

Next, we introduce a window function $W$ which is a $C^\infty$ scalar
field with
\beq
  W = 1+O(\rho^4/\calR^4) \text{ as } \rho \rightarrow 0,
\eeq
and $W \rightarrow 0$ sufficiently far from the particle, in particular in the
wavezone.
 Finally we define a regular remainder field
\beq
 \psi^R \equiv \psi^{\text{ret}} - W\tilde\psi^S
\eeq
which is a solution of
\beq
 \nabla^a\nabla_a\psi^R = - \nabla^a\nabla_a (W\tilde\psi^S)
             - 4\pi q \int_\gamma \delta^{(4)}(x-z(\tau)) d\tau
\label{eq:waveRdelta}
\eeq
from \eqn{eq:wave}.

\begin{figure}[h]
  \centering
  \includegraphics[width=9cm,angle=0]{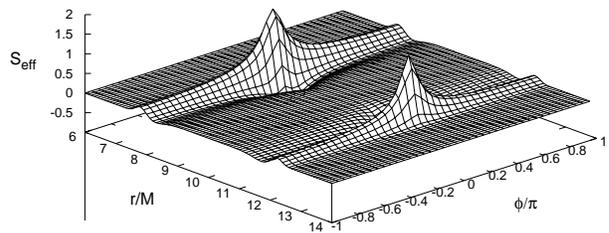}
  \caption{The effective source $S_{\text{eff}}$ on the equatorial plane. The particle is
  at $r/M=10$, $\phi/\pi=0$, where $S_{\text{eff}}$ appears to have no structure on this scale.
  The smooth ``double bump'' shape far from the charge is a characteristic of
  any function similar to $\nabla^2(W/|\vec r-\vec r_0|)$ in flat space,
  with a window function $W$ as given in \eqn{eq:window}. }
  \label{figS}
\end{figure}

\begin{figure}[h]
  \centering
  \includegraphics[width=9cm,angle=0]{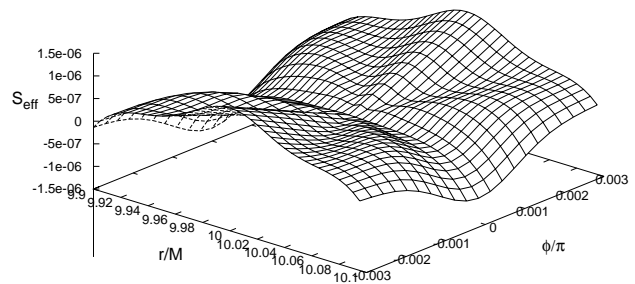}
  \caption{The effective source $S_{\text{eff}}$ in the equatorial plane
  in the vicinity of the point source at $r/M=10$, $\phi/\pi=0$. Note the significant difference
  of scales with Fig.~\ref{figS}.}
  \label{figSC0}
\end{figure}

The effective source of this equation
\beq
  S_{\text{eff}} \equiv - \nabla^a\nabla_a (W\tilde\psi^S)
             - 4\pi q \int_\gamma \delta^{(4)}(x-z(\tau)) d\tau
\label{eq:Seff}
\eeq
is straightforward to evaluate analytically, and the two terms on the
right hand side have delta-function pieces that precisely cancel at the
location of the charge, leaving a source which behaves as
\beq
  S_{\text{eff}} = O(\rho/\calR^4) \text{ as }
  \rho \rightarrow 0 .
\label{eq:SeffBehave}
\eeq
Thus the effective source $S_{\text{eff}}$ is continuous but not
necessarily  differentiable, $C^0$, at the particle while being $C^\infty$
elsewhere\footnote{With $\rho^2 \equiv x^2+y^2+z^2$, a function which is
$O(\rho^n)$ as $\rho\rightarrow0$, is at least $C^{n-1}$ where $\rho=0$.}.
Fig.~\ref{figS} shows the source function which is actually used in the
numerical analysis described in \S\ref{sec:finitedifference}. The modest
non-differentiability of $S_{\text{eff}}$ at the particle is revealed in
Fig.~\ref{figSC0}.

A solution $\psi^R$ of
\beq
  \nabla^a\nabla_a\psi^R = S_{\text{eff}}
\label{eq:waveR}
\eeq
is necessarily $C^2$ at the particle, and its derivative there provides
the self force acting on the particle. Also, in the wavezone $W$ effectively vanishes
and $\psi^R$ is then identically $\psi^{\text{ret}}$ and provides both the
waveform as well as any desired flux measured at a large distance.

General covariance dictates that the behavior of $S_{\text{eff}}$ in
\eqn{eq:Seff} may be analyzed in any coordinate system. But, only in the
specific coordinates of Refs.~\cite{ThorneHartle85} and \cite{Zhang86} is
it so easily shown \cite{DetweilerMW2003} that the simple expression for
$\psi^S$ in \eqn{eq:psiS} leads to the $O(\rho/\calR^4)$ behavior in
\eqn{eq:SeffBehave} and then to the $C^2$ nature of the solution $\psi^R$
of \eqn{eq:waveR}.

We describe the procedure of solving \eqn{eq:waveR} as \textit{field
regularization}. Then the derivatives of $\psi^R$ determine the
self-force, and $\psi^R$ is identical to $\psi^{\text{ret}}$ in the wave
zone.  With this process there is no apparent reason to determine the
actual retarded field. However, if one wants to compare results from field
regularization with results from a traditional determination of the
retarded field then simply adding $W\tilde\psi^S$ to the remainder
$\psi^R$ results in the retarded field $\psi^{\text{ret}}$.
  Such a comparison for our trial of field regularization appears in
Figs.~(\ref{TDFDcomp}) and (\ref{TDFDcompErr}).

\section{Numerical Implementation}
\label{sec:finitedifference}

As a concrete example and test of the field regularization prescription,
we apply it to the well-studied case of a scalar charge moving in a
circular orbit about a Schwarzschild black hole.  We choose $q/m = 1$ for
the charge to mass ratio of the particle, and a circular geodesic at
Schwarzschild radii, $R=10M$ and $R=12M$, where $M$ is the mass of the
black hole. We work in Schwarzschild coordinates, in which the metric is
expressed as $g_{ab} = \mbox{diag}(-(1-2M/r),\;
1/(1-2M/r),\;r^2,\;r^2\sin^2{\theta})$.

For this task, we have developed code that (a) solves the
regularized wave equation (\ref{eq:waveR}) and (b) computes the scalar
self-force. For simplicity, we have chosen to solve the regularized wave
equation using a (1+1)-approach. We exploit the spherical symmetry of the
background, decompose physical quantities into spherical harmonics, and
then solve the resulting set of (1+1)D-wave equations (one `time' + one `space') for the
spherical-harmonic components.

It must be stressed at this point that
the numerical implementation presented in this paper does not highlight
the advantages of our prescription. The simplicity of the orbit we consider
and the spherical symmetry of our background geometry naturally lend
themselves to a significantly more efficient frequency-domain approach.
But the point here is to provide a quick, first check of our ideas. One is cautioned
not to let the simplicity of the present problem obscure the
generality of our proposed method, and its potential for cases with
generic orbits and spacetimes lacking symmetry, and for self-consistent evolutions
which are likely to require self-force calculations in real time (as opposed to being a
post-processing step). Testing the robustness of our method against these more
difficult problems will be addressed in future work. The current goal
is mainly to establish plausibility: to provide both an initial proof-of-principle for the method
and also the necessary practice en route to tackling more interesting
problems handled using more sophisticated numerical techniques.

\subsection{Scalar fields in a Schwarzschild geometry}

Wave equations in spherically-symmetric backgrounds simplify considerably with a spherical-harmonic
decomposition of the field. In the case of a Schwarzschild geometry expressed in Schwarzschild coordinates, this
decomposition is typically performed as follows:
\begin{equation}
\psi = \sum_{lm} \frac{1}{r}f_{lm}(r)(t,r)Y_{lm}(\theta,\phi).
\end{equation} With $r_*=r+2M\ln\left(r/2M-1\right)$, this yields equations for $f_{lm}(t,r_*)$:
\begin{equation}
-\frac{\partial^2 f_{lm}}{\partial t^2} + \frac{\partial^2 f_{lm}}{\partial r_*^2} - V(r_*)f_{lm} =
S_{lm}(t,r_*) \label{eq:lmwave}
\end{equation} where $V(r_*)$ is implicitly given in terms of $r$ as:
\begin{equation}
V(r_*) = \left(1-\frac{2M}{r}\right)\left[\frac{l(l+1)}{r^2}+\frac{2M}{r^3}\right],
\end{equation} while the source $S_{lm}(t,r_*)$ is
\begin{equation}
S_{lm}(t,r_*) = (r-2M)\int \rho(x_{\alpha}')Y_{lm}(\theta',\phi') d\Omega'.
\end{equation}
In a frequency-domain approach, one further chooses to Fourier-decompose
$f_{lm}(t,r_*)=\int F_{lm\omega}(r_*) \exp(-i\omega t)d\omega$, and thereby solve
the resulting set of ordinary differential equations for
$F_{lm\omega}(r_*)$, for each mode $\omega$. This method tends to be
numerically expensive, however, for sources with a continuous
$\omega$-spectrum. Instead, we choose to solve Eq.~(\ref{eq:lmwave}) as an
initial boundary value problem, in a time-domain fashion, for each
$(l,m)$. This is done with $S_{lm}$ computed beforehand as the spherical
harmonic components of the effective source found  in Eq.~(\ref{eq:Seff}).

\subsection{Effective source term}

A novel feature of our approach is the use of an effective source that permits the easy
calculation of both self-forces and fluxes. As discussed above, this effective source is formally
\begin{equation}
S_{\tiny\mbox{eff}}=-\nabla^2(W\tilde\psi^S)-4\pi q \int_\gamma \delta^{(4)}(x-z(\tau)) d\tau .
\end{equation}

To lowest order, the singular field takes on the form
\begin{equation}
\psi^S \approx \tilde\psi^S = \frac{q}{\rho}.
\end{equation}

We take advantage of the results in \cite{DetweilerMW2003}, where $\rho$
is expressed explicitly as $\rho=\sqrt{\eta_{ij}x^ix^j}$ in
Thorne-Hartle-Zhang coordinates for a particle moving in a circular orbit.
Using the coordinate transformation found in Appendix B of
\cite{DetweilerMW2003}, where a more detailed discussion of the singular
field is found, we are able to express the singular field in Schwarzschild
coordinates. (A brief discussion of this coordinate transformation is
provided in Appendix A). To complete our effective source, we select a
window function whose role is to kill off smoothly the singular field in
regions where it is not needed. Consequently, the effective support of the
windowed singular field $W\tilde\psi^S$ is confined to a compact region
surrounding the particle's world line.

Our chosen window function is spherically-symmetric with respect to the
center of the black hole. This choice was not necessary but guarantees
that $W$ would not unnecessarily modify the $(l,m)$-spectrum of the
source, and thereby allows us to make more controlled comparisons with
existing frequency-domain results on the same problem. Our simple choice
of $W$ is
\begin{align}
W(r) = \exp\Big[-\frac{(r-R)^N}{\sigma^N}\Big].
\label{eq:window}
\end{align}
In this window function, the constant $\sigma$ sets the width, and the
exponent $N$ controls how quickly $W$ and $\nabla_a W$ reach the required
values of 1 and 0, respectively, as one approaches the particle. We use
$\sigma=2M$ and $N=8$ in all the results presented in this paper. It is
necessary that $N$ is an even integer, and taking full advantage of the
accuracy of our approximation for $\psi^S$ requires that $W = 1 +
O(\rho^4/\calR^4)$ as $\rho\rightarrow0$. Thus we require that $N\ge4$. In
fact we used $N=8$ in anticipation of improving the approximation for
$\psi^S$ in the future.

Our choice for the window function leads to the effective source
$S_{\text{eff}}$ displayed in Figs.~(\ref{figS}) and (\ref{figSC0}). A
larger choice for $\sigma$ would spread the bumps out further, and a
smaller choice for $N$ would smooth the bumps. But if $N$ were less than
4, then $W\tilde\psi^S$ would not adequately match the behavior of
$\psi^S$ as $\rho\rightarrow0$.

With the effective source constructed as above, its spherical-harmonic
components were then computed. Circular orbits proved advantageous here
because of which the time dependence of the components could then
simply be inferred. The spherical harmonic components were
evaluated with a 4th-order Runge-Kutta integrator with self-adjusting step
size, which was derived from a routine in \cite{NumericalRecipes}.

\subsection{Evolution algorithm}

The integration scheme we use in evolving Eq.~(\ref{eq:lmwave}) follows a
technique first introduced by Lousto and Price \cite{LoustoP1997}, and
later improved to fourth-order accuracy by Lousto \cite{Lousto2005} and
Haas \cite{Haas2007}. Unlike their schemes, however, we do not deal with
sourced and vacuum regions of our numerical domain separately. Their use
of a singular delta-function source meant that the resulting field was
non-differentiable at the location of the charge, while smooth everywhere
else. For us, the effective source is ${C}^0$, implying that the field is
at least ${C}^2$. While this is still of finite differentiability, we find
that the effective source is differentiable enough not to warrant a
treatment different from the vacuum case. \vspace{1em}

In the $(t,r_*)$-plane, we introduce a staggered grid with step sizes $\Delta t = \frac{1}{2} \Delta r_* = h$.
In this grid, a unit cell is defined to be the diamond region with corners
$\{(t+h,r_*),(t-h,r_*),(t,r_*+h),(t,r_*-h)\}$. Only at these grid points do we evaluate $f_{lm}$. We henceforth
drop the spherical-harmonic indices in $f_{lm}$ for convenience.

\begin{figure}[h]
  \centering
  \includegraphics[width=7.5cm,angle=0]{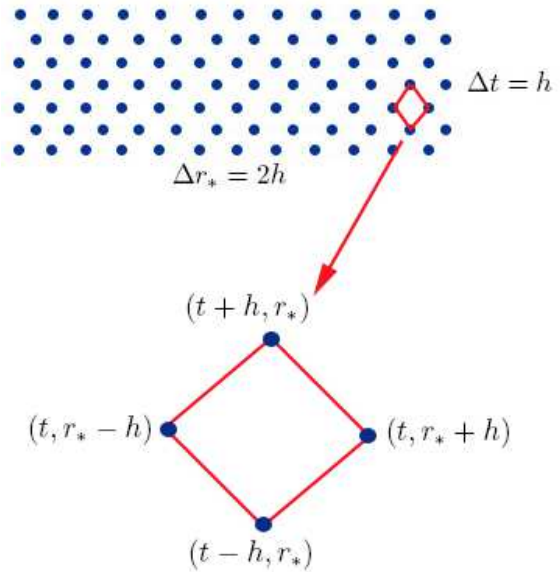}
  \caption{Staggered (characteristic) grid with unit cell.}
 \label{grid1}
\end{figure}

The main idea behind the algorithm is to integrate the wave equation over
a unit cell. This is done easiest with Eddington-Finkelstein null
coordinates $u=t-r_*$ and $v=t+r_*$ as the integration variables.

The differential operator of the wave equation, when expressed in
$(u,v)$ coordinates, is just $-4\partial_u\partial_v$. Over a unit cell
then, the derivative term in Eq.~(\ref{eq:lmwave}) can be integrated
exactly:
\begin{align}
    \iint_C \,-4\,\partial_u\partial_v\,f \,du\,dv &=-4[f(t+h,r_*)+f(t-h,r_*)\nonumber\\&
                                                    -f(t,r_*+h)-f(t,r_*-h)]. \label{eq:intwave}
\end{align}

Integrations of the potential term and the source term do not enjoy the
same simplicity as the derivative term. We need to approximate these
integrals to the appropriate order in $h$ so as to achieve the desired
${O}(h^4)$-convergence over the entire numerical domain.

Suppose we wish to solve the wave equation over a region defined by
$\Delta T$ and $\Delta R_*$. In this region, there will be $N=\Delta
T\Delta R_*/h^2$ cells. Achieving ${O}(h^4)$-convergence for evolution
means that we need to integrate the wave equation with an over-all error
of at most ${O}(h^4)$ over the entire computational domain. For a unit cell,
this means an approximation with an error ${O}(h^4)/N \sim {O}(h^6)$.

Such an approximation is achieved with the double Simpson rule. Consider a sufficiently differentiable function
$G(t,r_*)$ to be integrated over a unit cell. The double Simpson rule then reads:
\begin{align}
\iint_C& G \,du\,dv = \left(\frac{h}{3}\right)^2 [G_{\tiny\mbox{corners}}+16G(t,r_*)\nonumber\\
                   & +4(G(t+h/2,r_*-h/2)
                   +G(t+h/2,r_*+h/2)\nonumber\\
                   &+G(t-h/2,r_*-h/2)
                   +G(t-h/2,r_*+h/2))] \nonumber\\&+{O}(h^6), \label{eq:simpson}
\end{align} where $G_{\tiny\mbox{corners}}$ is just the sum of the values of $G$ evaluated at the
corners of the unit cell.

This is directly applied in integrating the source term of Eq.~(\ref{eq:lmwave}):
\begin{equation}
  \iint_C S^{\tiny \mbox{eff}}_{lm} \,du\,dv.
\end{equation}
One simply evaluates the source term at the required points and then sums
these accordingly in order to get an ${O}(h^6)$-accurate approximation to
the integral.

However, for integrating the potential term:
\begin{equation}
\iint_C -Vf \,du\,dv, \label{eq:intpot1}
\end{equation}
we recall that one has only restricted access to $f$. The direct evaluation of $f$ is done only at the grid points,
i.e. corners of the unit cell. Thus far, only $G_{\tiny\mbox{corners}}$  in Eq.~(\ref{eq:simpson}) can
be explicitly evaluated. To use Eq.~(\ref{eq:simpson}) for the potential term, we need to determine
how to evaluate $f$ at all the other points.

Following Lousto \cite{Lousto2005}, we evaluate $G=-Vf$ at the central grid point (i.e. $G(t,r_*)$) using values at
the neighboring grid points on the same time slice.
\begin{align}
G(t,r_*) = &\frac{1}{16}[9G(t,r_*-h)+9G(t,r_*+h)\nonumber \\ &-G(t,r_*-3h)-G(t,r_*+3h)] +
{O}(h^4).\label{eq:center}
\end{align} Note that this is different from Haas \cite{Haas2007}, who uses grid points in the causal past
of the unit cell. The ${O}(h^4)$-error incurred in this approximation is
tolerable because of the $h^2$-factor that appears in
Eq.~(\ref{eq:simpson}).

We seek similar approximations for $G$ in the remaining points. 
Consider first the pair $G(t+h/2,r_*-h/2)$ and $G(t-h/2,r_*-h/2)$. (The other pair, composed of $G(t+h/2,r_*+h/2)$ and $G(t-h/2,r_*+h/2)$, is treated similarly). This pair makes up the top and bottom corners of a smaller cell, ${\mathcal C}_{\tiny \mbox{left}}$ , made up of the points $\{(t+h/2,r_*-h/2),(t-h/2,r_*-h/2),(t,r_*-h),(t,r_*)\}$.\vspace{1.5em}

\begin{figure}[t]
  \centering
  \includegraphics[width=8.5cm,angle=0]{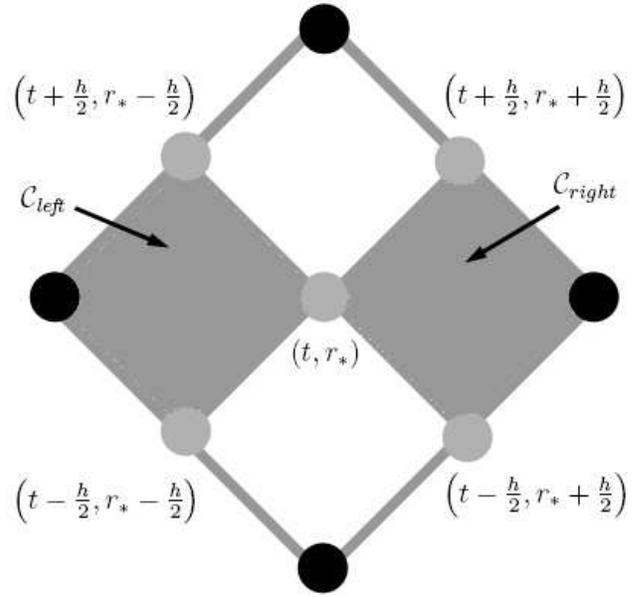}
  \caption{Unit cell of the algorithm. The black dots indicate grid points, whereas the gray ones stand for the points where
           $G=-Vf$ needs to be approximated. The subcells ${\mathcal C}_{le\!f\!t}$ and ${\mathcal C}_{right}$ are shaded
           gray. To approximate $G$ at some of the gray dots, we integrate the wave equation in each of these subcells.}
 \label{figschwrt}
\end{figure}

What we shall do next is find an approximation for
\begin{align}
G(t+h/2,r_*-h/2)+ G(t-h/2,r_*-h/2)
\end{align}
accurate to ${O}(h^4)$. Again, this is sufficient because of the
$h^2$-factor in Eq.~(\ref{eq:simpson}).

Consider integrating the wave equation over this smaller cell, but this
time only up to an accuracy of ${O}(h^4)$. The integral over the
derivative term will again be exact:
\begin{align}
    \iint_{C_{\tiny \mbox{left}}} &\!\!\!-4\,\partial_u\partial_v\,f \,du\,dv =-4[f(t+h/2,r_*-h/2)\nonumber\\&+f(t-h/2,r_*-h/2)
                                                    -f(t,r_*-h)-f(t,r_*)]. \label{eq:intderiv}
\end{align}
The integrals of the potential and source terms over this smaller cell are
again handled as before, but this time we approximate them only to
${O}(h^4)$. To this end, the double trapezoidal rule will suffice, which
reads:
\begin{align}
\iint_{C_{\tiny \mbox{left}}} & G \,du\,dv = \left(\frac{h}{2}\right)^2 [G(t+h/2,r_*-h/2)\nonumber\\
                                        + &G(t-h/2,r_*-h/2)
                                        + G(t,r_*-h)+G(t,r_*)]
                                        \nonumber \\+ &{O}(h^4). \label{eq:trap}
\end{align}
Applying this to the potential term then gives:
\begin{align}
\iint_{C_{\tiny \mbox{left}}} -Vf \,du\,dv = &-\left(\frac{h}{2}\right)^2\times \nonumber \\
                        &[V(r_*-h/2)f(t+h/2,r_*-h/2) \nonumber \\ &+V(r_*-h/2)f(t-h/2,r_*-h/2) \nonumber\\
                                        &+V(r_*-h)f(t,r_*-h) \nonumber\\&+V(r_*)f(t,r_*)]
                                        +{O}(h^4). \label{eq:intpot2}
\end{align}
Combining Eq.~(\ref{eq:intderiv}) and Eq.~(\ref{eq:intpot2}), the result of integrating the wave equation over this
smaller cell yields:
\begin{align}
f&(t+h/2,r_*-h/2)+f(t-h/2,r_*-h/2) = \nonumber
\\&\left(f(t,r_*-h)+f(t,r_*)\right)
                     \left[1-\frac{1}{2}\left(\frac{h}{2}\right)^2V(r_*-h/2) \right]
                    \nonumber \\ &-\frac{1}{4}\iint_{C_{\tiny \mbox{left}}} S_{\tiny \mbox{eff}}\,\,du\,dv+ {O}(h^4).
\end{align}

After multiplying both sides of this last equation by $-V(r_*-h/2)$, the
resulting left-hand-side becomes two of the
as yet missing pieces
in the double Simpson formula:
\begin{align}
G(t+h/2,r_*&-h/2)+ G(t-h/2,r_*-h/2) = \nonumber \\ &- V(r_*-h/2)f(t+h/2,r_*-h/2)
                                    \nonumber \\ &-V(r_*-h/2)f(t-h/2,r_*-h/2).
\end{align}
The resulting equation then gives us the desired ${O}(h^4)$-approximation of
the missing
expression, $G(t+h/2,r_*-h/2)+ G(t-h/2,r_*-h/2)$, in Eq.~(\ref{eq:simpson}).
Following the same steps, it is easy to arrive at an equivalent approximation
for the other missing
pair, $G(t+h/2,r_*+h/2)+ G(t-h/2,r_*+h/2)$. We summarize these below:
\begin{align}
G(t+h/2,&\, r_*-h/2)+G(t-h/2,r_*-h/2) = \nonumber \\&-V(r_*-h/2)\left(f(t,r_*-h)+f(t,r_*)\right)
                     \nonumber \\&\times\left[1-\frac{1}{2}\left(\frac{h}{2}\right)^2V(r_*-h/2) \right]
                    \nonumber \\&+\frac{V(r_*-h/2)}{4}\iint_{C_{\tiny \mbox{left}}} S_{\tiny \mbox{eff}}\,\,du\,dv+ {O}(h^4).
                    \label{eq:LH1}
\end{align}
\begin{align}
G(t+h/2,&\, r_*+h/2)+G(t-h/2,r_*+h/2) = \nonumber \\&-V(r_*+h/2)\left(f(t,r_*+h)+f(t,r_*)\right)
                     \nonumber \\&\times\left[1-\frac{1}{2}\left(\frac{h}{2}\right)^2V(r_*-h/2) \right]
                    \nonumber \\&+\frac{V(r_*+h/2)}{4}\iint_{C_{\tiny \mbox{right}}} S_{\tiny \mbox{eff}}\,\,du\,dv+ {O}(h^4).
                    \label{eq:LH2}
\end{align} Except for the presence of integrated source terms, these equations are identical to Lousto's equations (32) and (33),
and Haas's equations (2.8) and (2.9).
\begin{widetext}
Following Haas \cite{Haas2007}, we choose to avoid isolated occurrences of
$f(t,r_*)$, which prove to be numerically unstable close to the event horizon. As pointed out in \cite{Haas2007}, this is
due to having first approximated $G=-Vf$, which makes it difficult to isolate $f=-G/V$ where $V\approx0$. This appears
unnecessary if $f$ were directly approximated instead of $G$ in (\ref{eq:center}). Nevertheless, like Haas, we avoid needing to isolate
$f$ by adding up equations Eq.~(\ref{eq:LH1}) and Eq.~(\ref{eq:LH2}), and then Taylor-expanding the potential terms that
are multiplied by $f(t,r_*)$.  The result is Haas's equation (2.10) with extra source terms:
\begin{align}
\sum G \equiv G(t+h/2,&r_*-h/2)+G(t-h/2,r_*-h/2)+G(t+h/2,r_*+h/2)+G(t-h/2,r_*+h/2)= \nonumber \\
              &-2V(r_*)f(t,r_*)\left[1-\frac{1}{2}\left(\frac{h}{2}\right)^2\right]
              -V(r_*-h/2)f(t,r_*-h)\left[1-\frac{1}{2}\left(\frac{h}{2}\right)^2V(r_*-h/2)\right]\nonumber \\
              &-V(r_*+h/2)f(t,r_*+h)\left[1-\frac{1}{2}\left(\frac{h}{2}\right)^2V(r_*+h/2)\right]\nonumber \\
              &-\frac{1}{2}[V(r_*-h/2)-2V(r_*)+V(r_*+h/2)](f(t,r_*-h)+f(t,r_*+h)) \nonumber \\
              & +\frac{V(r_*-h/2)}{4}\iint_{C_{\tiny \mbox{left}}} S_{\tiny \mbox{eff}}\,\,du\,dv+\frac{V(r_*+h/2)}{4}\iint_{C_{\tiny \mbox{right}}} S_{\tiny \mbox{eff}}\,\,du\,dv
              + {O}(h^4) \label{eq:smallcells}
\end{align}

This last equation completes the pieces needed for the evolution algorithm.

Using Eq.~(\ref{eq:intwave}) for the derivative term and Eq.~(\ref{eq:simpson}) for the potential and source terms, the
result of integrating the wave equation over the unit cell finally yields:
\begin{align}
f(t+h,r_*)=&-f(t-h,r_*)
+\frac{\left[1-\frac{1}{4}(\frac{h}{3})^2V(r_*+h)\right]}{\left[1+\frac{1}{4}(\frac{h}{3})^2V(r_*)\right]}f(t,r_*+h)
+\frac{\left[1-\frac{1}{4}(\frac{h}{3})^2V(r_*-h)\right]}{\left[1+\frac{1}{4}(\frac{h}{3})^2V(r_*)\right]}f(t,r_*-h)
\nonumber \\&-\frac{1}{\left[1+\frac{1}{4}(\frac{h}{3})^2V(r_*)\right]}\left[\left(\frac{h}{3}\right)^2\left(4G_0+ \sum G\right) +\frac{1}{4}\iint_{C}
S_{\tiny \mbox{eff}}\,\,du\,dv\right]+{O}(h^6), \label{eq:fullscheme}
\end{align}
\end{widetext}
where $G_0$ is evaluated according to Eq.~(\ref{eq:center}), with $G(t,r_*)=-V(r_*)f(t,r_*)$; $\sum G$ is the
expression in Eq.~(\ref{eq:smallcells}); and the double Simpson rule Eq.~(\ref{eq:simpson}) is applied in evaluating
the remaining integral term $\iint_C S_{\tiny \mbox{eff}}\, dudv$. With this equation, one can now determine the field
$f$ at time $t+h$ given its values at earlier times $t$ and $t-h$.

This derivation makes liberal use of double Simpson and double trapezoidal formulas when approximating
integrals of the source and potential terms over the unit cell. The formulas come
from their single-integral counterparts:
\begin{align}
\int^{x_0+h}_{x_0} f(x) dx = \frac{h}{2}\left[f(x_0)+f(x_0+h)\right]-\frac{h^3}{12}f^{(2)}(\xi)
\end{align}
\begin{align}
\int^{x_0+2h}_{x_0} f(x) dx = \frac{h}{3}&\left[f(x_0)+4f(x_0+h)+f(x_0+2h)\right]\nonumber
\\&-\frac{h^5}{90}f^{(4)}(\xi),
\end{align}
where $f^{(n)}$ denotes the $n$th-derivative of $f$, and $\xi$ is some point
within the limits of integration. These require the boundedness, if not
existence of the second and fourth derivatives of the integrand for the error
estimate to be valid. With the limited differentiability of our source and
potential terms (${C}^0$ and ${C}^2$, respectively), one might worry about the validity of
our over-all convergence estimate. However, our calculations reveal that 4th-order
convergence is achieved despite this deficiency.

\subsection{Initial data and boundary conditions}

For our evolution we have no obvious method for choosing a priori the
correct initial data, which consists of the value of $\psi^R$ on two
consecutive constant-time slices. Consequently we just set the initial
$\psi^R$ to zero everywhere on the initial two slices. Physically, this
scenario corresponds to the impulsive appearance of the scalar point
charge along with $W\tilde\Psi^S$, which leads to spurious radiation
contaminating our computational domain during the early stages of the
evolution. Fortunately, this radiation propagates out of the regions of
interest quickly; so to circumvent the need for proper initial data, we
simply evolve the equation to long enough times such that initial data
effects do not become pertinent in any of our results.

With the scalar charge moving in a circular orbit, it is expected that the
field eventually becomes stationary in a frame corotating with the charge. A
practical test then for the persistence of initial data effects is to simply
check whether or not the field has already settled into a quiescent state
when evaluated in this frame.

Boundary conditions are treated similarly. Rather than handling them carefully, we instead made the
computational domain large enough that errors incurred by unspecified boundary conditions did not affect our
regions of interest. For this work, our choice of boundaries were at $r_*=-700M$ and $r_*=800M$.

\subsection{Self-force calculation}

At the end of evolution for each mode, we compute the self-force at the location of the particle.
Since this location is not on any grid point, interpolation of $f_{lm}(T,r)$ and its derivatives
to $r=R$  was required using a selection of grid points surrounding it.

Once this was done, computing the self-force was a simple matter of performing the following sums:
\begin{align}
\psi^R = &\frac{1}{R}\sum_{l=0}^Lf_{00}(T,R)Y_{00}\left(\frac{\pi}{2},\Omega T \right) \nonumber \\&+\frac{2}{R}\sum_{l=1}^L\sum_{m=1}^l \mbox{Re}\left(f_{lm}(T,R)Y_{lm}\left(\frac{\pi}{2},\Omega T\right)\right)
\end{align}
\begin{align}
\partial_t\psi^R = \frac{2}{R}\sum_{l=0}^L\sum_{m=1}^l (m\Omega)\mbox{Im}\left(f_{lm}(T,R)Y_{lm}\left(\frac{\pi}{2},\Omega T\right)\right)
\end{align}
\begin{align}
\partial_r\psi^R = &\frac{1}{R}\sum_{l=0}^L\partial_rf_{00}(T,R)Y_{00}\left(\frac{\pi}{2},\Omega T \right) \nonumber \\&
+\frac{2}{R}\sum_{l=1}^L\sum_{m=1}^l \mbox{Re}\left(\partial_rf_{lm}(T,R)Y_{lm}\left(\frac{\pi}{2},\Omega T\right)\right)
\end{align}
Here, $L$ is the point where we truncate the multipole expansion. In all our work we have used $L=39$. These sums arise primarily
because our charge moves in a circular orbit.

Two methods were employed for interpolation. The first was a simple Lagrange interpolation of both
$f$ and $\partial_rf$ to $r=R$. However, because of the finite differentiability of our regular field at $r=R$
we also interpolated using the form
\begin{align}
\psi^R(r) = A_0 + A_1x+A_2x^2+A_3x^3
    + \theta(x)B_0x^3,
\end{align}
where $x=r-R$, and $\theta(x)$ is the standard Heaviside function. This
form closely respects the ${C}^2$ nature
 of the regular field at $r=R$ by allowing for a discontinuity in the third derivative.

With this form, $(\nabla_rF)|_{r=R}=A_1$. However, this led to results not significantly different from the one achieved with
ordinary Lagrange interpolation.

\section{CODE DIAGNOSTICS}
\label{sec:diagnostics}

\subsection{Convergence}

The convergence of a time-domain code is easily determined by computing
the convergence factor $n$ as defined by Lousto\cite{Lousto2005}:
\begin{align}
n(r_*,t) =& \log \left|\frac{f_{4h}(r_*,t)-f_{2h}(r_*,t)}{f_{2h}(r_*,t)-f_{h}(r_*,t)}\right|/\log(2) \nonumber \\ &+ \log|\epsilon^{(n)}(\xi)|/\log(2),
\end{align}
where $f_{\Delta}(r_*,t)$ is the result of the evolution for a resolution of $\Delta$, and $\epsilon^{(n)}(\xi)$ represents
an error function $\approx 1$. An $n$th-order evolution code is one for which $\psi = \psi_N(h) +(\epsilon^{(n)})(\xi)h^n$ ,
where $\psi_N(h)$ is the numerical solution at resolution $h$.

In checking convergence, one evolves the wave equation at different
resolutions, $h$, $2h$, and $4h$. For a fixed $r_*=R$, one then extracts
$f_{h}(R,t), f_{2h}(R,t),$ and $f_{4h}(R,t)$ for all $t$. From these, one can
compute $n(R,t)$.

\begin{figure}[h]
  \centering
  \includegraphics[width=9cm,angle=0]{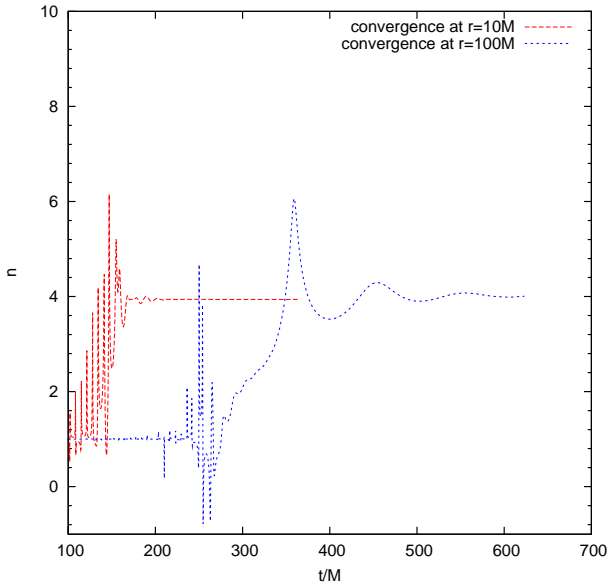}
  \caption{Convergence at the particle location ($r=10M$) and in the wavezone ($r=100M$). At the start of the evolution, inequivalent initial data
  lead to the lack of 4th-order convergence. But $n$ gradually approaches 4 as initial-data effects propagate away from the computational domain.
  Note that the convergence test at the particle location already includes the interpolation step. }
  \label{convergence}
\end{figure}

The convergence factor was computed for a few representative points in the
wavezone and in the region close to the point particle. Two of these are
shown in Fig.~\ref{convergence}. These are for $r\approx 10M$ and $r\approx
100M$. All show the desired 4th-order convergence eventually, following a
transient period in which the numerical evolution is contaminated by the
effects of poor initial data.

\subsection{High-$l$ fall-off}

In \cite{BarackO2000, Lousto00, BMNOS02, DetweilerMW2003}, it was
demonstrated that the rate of convergence of the $l$-components of the
self-force was dictated primarily by the lack of differentiability of the
regular piece from which the self-force is computed. By definition, the
difference between the retarded field and the singular field yields a
function that is ${C}^\infty$. In this ideal situation, convergence in $l$
of the self-force computed from this smooth regular field would be
exponentially fast. In practice, however, one is always limited to
constructing only an approximate singular field, therefore leaving
non-differentiable pieces in the residual $\psi^{\text{ret}} -
\tilde\psi^S$. The degree of non-differentiability of this remainder is
what sets the rate of convergence of the self-force in $l$.

The high-$l$ asymptotic structure of the singular piece $\psi^S$ is such that:
\begin{align}
\mathop{\lim }\limits_{r \to R}(\nabla_r&\psi^S)_l = \left(l+\frac{1}{2}\right)A_r+B_r-\frac{2\sqrt{2}D_r}{(2l-1)(2l+3)}\nonumber \\
&+\frac{E^{(1)}_r\mathcal{P}_{3/2}}{(2l-3)(2l-1)(2l+3)(2l+5)} + \ldots, \label{regularization}
\end{align}
where $A,B,D,\ldots$ are the \emph{regularization parameters}, which
commonly appear in contemporary self-force studies \cite{BarackO2000,
BMNOS02}.

The number of regularization parameters that can be determined in this expansion
corresponds directly to the accuracy of the singular field approximation. Convergence in
$l$ of the self-force $\nabla_r\psi^R$ is then fixed by the lowest-order
undetermined piece of the approximate singular field. Specifically, if the
singular field is accurately determined only up to the $B$-term of the
expansion above, then the $l$-convergence of the self-force would be $\sim
1/l^2$, corresponding to the $D$-term fall-off.

\begin{figure}[h]
  \centering
  \includegraphics[width=9cm,angle=0]{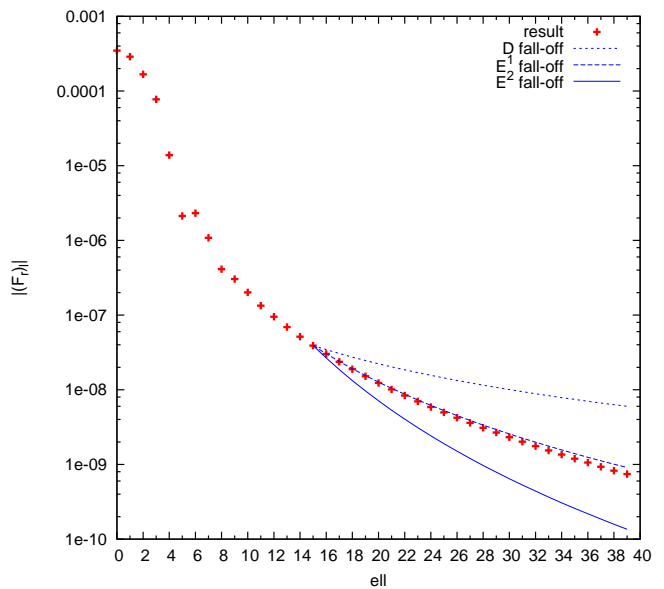}
\caption{$(\nabla_r\psi^R)_l$ versus $l$. Our results show $l$-convergence
closest to the $E^{(1)}$ fall-off. The blue lines correspond to the expected
fall-off in the $r$-component of the self-force when one regularizes using
a singular field approximation without the $D$-term, $E^{(1)}$-term, and $E^{(2)}$-term,
respectively. Our result is matched to these curves at $l$=15.}
  \label{highL}
\end{figure}

The approximation to the singular field here is $\tilde\psi^S = q/\rho$, and
the attendant THZ-Schwarzschild coordinate transformation, has been shown in
\cite{DetweilerMW2003} to include
at least the $D$-term. The expectation then
would be for the $l$-components of our remainder, $(\nabla_r\psi^R)_l$, to
fall off as the $E^{(1)}$-piece:
\begin{align}
\frac{E^{(1)}_r\mathcal{P}_{3/2}}{(2l-3)(2l-1)(2l+3)(2l+5)}.
\end{align}
Fig.~\ref{highL} shows our results confirming this expectation. Our
results are plotted with $D$, $E^{(1)}$ and $E^{(2)}$ fall-off curves
found in Eq.~(\ref{regularization}) that are made to match at $l=15$.

\begin{figure}[h]
  \centering
  \includegraphics[width=9cm,angle=0]{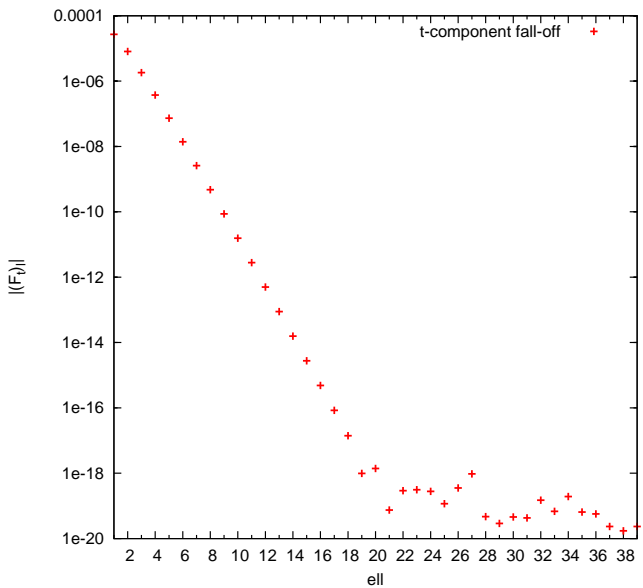}
  \caption{$(\nabla_t\psi^R)_l$ versus $l$. The expected exponential fall-off is observed until the point where numerical noise begins to dominate.}
 \label{t-falloff}
\end{figure}

The $t$-component of the self-force, on the other hand, does not require
regularization for the case of a charge in a circular orbit of Schwarzschild.
An exponential fall-off is then expected. This is shown in
Fig.~\ref{t-falloff}.

\subsection{Dependence on the window function}

The use of a window function $W$ is a peculiar feature of our approach. Its
function is mainly to kill off $\tilde\psi^S$ in the regions where it is no
longer relevant and thereby to have the computed regular field $\psi^R$
transform into the retarded field in those regions. As it is a mere
artifact of our implementation, it is crucial that the self-force and
waveform be independent of the specific choice of window function.

One has considerable freedom in choosing $W$, the only requirements being
that $W$ goes to 1 and that its gradient vanishes fast enough in the limit
that one approaches the point particle. With our specific choice of W
becoming numerically significant only in an annular region $|r-R| \lesssim
\sigma$, we have inspected the changes in the self-force and fluxes as one
varies the width $\sigma$.

\begin{figure}[h]
  \centering
  \includegraphics[width=9cm,angle=0]{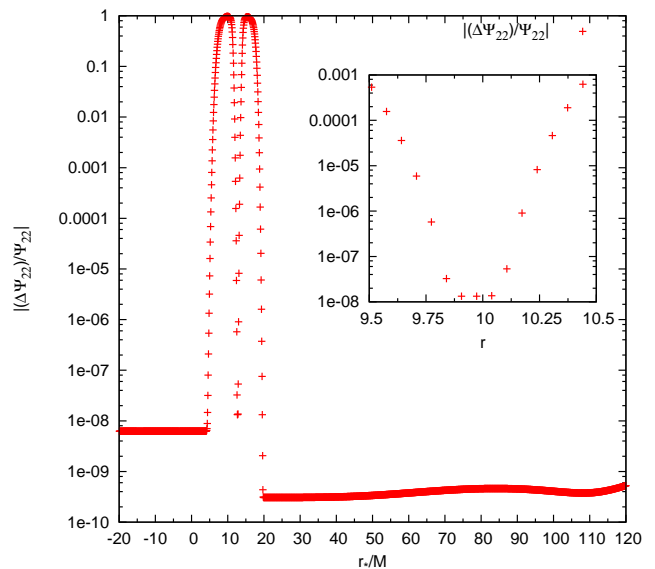}
  \caption{Fractional changes in \textbf{$\psi^R_{22}$} as a result of using different window functions.
  Note that these changes are significant only where the
  window functions differ; they are insignificant
  in the important regions in the vicinity of the charge, $r=10M$, and in the
  wave zone, $r\gg 10M$.}
 \label{windowcomp}
\end{figure}

Fig.~\ref{windowcomp} shows the effect of doubling the annular support of
the window function. The $(l=2,m=2)$ wave equation was evolved for the
same length of time, but with effective sources having different window
functions. A comparison is then made of the resulting fields over most of
the computational domain. It is seen that the fields differ significantly
only in regions where the window functions differ. Nevertheless, the
regular field $\psi^R$ remains the same (up to fractional changes of $\sim
10^{-8}$ in the most physically-relevant regions: the vicinity of the
charge, $r=10M$ (where the self-force is computed), and the wavezone,
$r\gg 10M$ (where the waveform is to be extracted).

As desired then, the window function appears to have no effect on any of the numerical results attained.

\section{Results}
\label{sec:results}

\subsection{Recovering the retarded field}

\begin{figure}[h]
  \centering
  \includegraphics[width=9cm,angle=0]{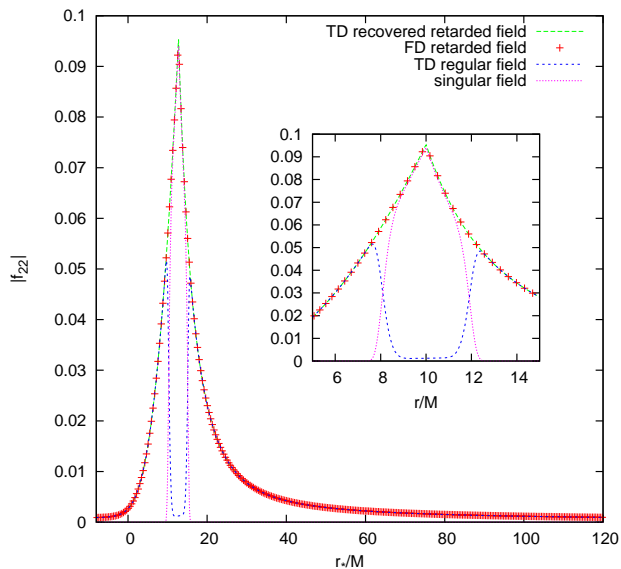}
  \caption{Comparison of time-domain and frequency-domain results for $f_{22}(r_*)$. The regular field is the result of our code (represented by the blue dashed line). Adding this to
            the ($l$=2,$m$=2)-component of our analytical singular field, $W\tilde\psi^S$, results in the FD-computed retarded field to good agreement.}
 \label{TDFDcomp}
\end{figure}

\begin{figure}[h]
  \centering
  \includegraphics[width=9cm,angle=0]{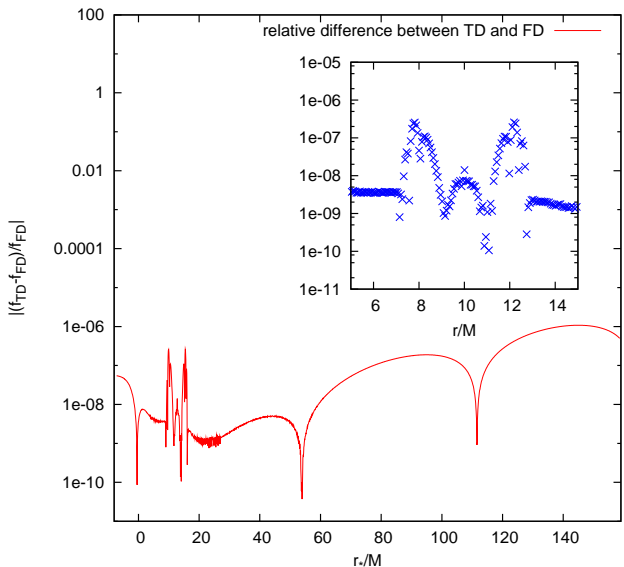}
  \caption{Relative error between time-domain and frequency-domain results for $f_{22}(r_*)$. Excellent agreement is achieved;
  errors are at worst $\sim 10^{-6}$.}
 \label{TDFDcompErr}
\end{figure}

From our numerical calculations we are able to accurately recover the
retarded field. In the wavezone, where the singular field is negligible,
this retarded field equals our regular field $\psi^R$. Since, energy
fluxes depend directly on the retarded field in this region, the accuracy
with which we recover the retarded field in the wavezone gives us a
measure of how well we can compute fluxes using our method. We determine
this accuracy by comparing our result for $\psi^R$ in the wavezone with
that obtained for the retarded field using a separate frequency-domain
calculation. An example of such a comparison is shown in
Figs.~\ref{TDFDcomp} and \ref{TDFDcompErr}. We observe relative errors
that are at worst $10^{-6}$. Shown in Fig.~\ref{TDFDcomp} are the
$(l=2,m=2)$ component of the retarded field computed in the
frequency-domain and our corresponding time-domain result,
$\psi_{22}^R+(W\tilde\psi^S)_{22}$, for the case of a charge at $r=10M$.
Also shown are the singular field $(W\tilde\psi^S)_{22}$ and the regular
field $\psi_{22}^R$.

\subsection{Self-force}

We obtain the $t$ and $r$ components of the self-force for an orbit at
radii $R=10M$ and $12M$. These are summarized in Table~\ref{tab:only}.

\begin{figure}[h]
  \centering
  \includegraphics[width=9cm,angle=0]{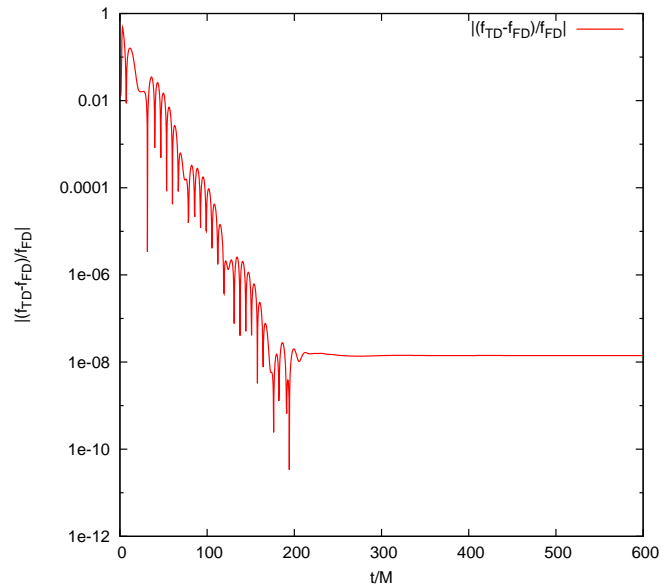}
  \caption{ 
  Relative error in the time-domain calculation of $f_{22}$, as
  compared with the frequency-domain calculation, versus time at $r$ close to the charge at
  $10M$.}
 \label{equilibrium}
\end{figure}

Fig.~\ref{equilibrium} shows the convergence of our calculation of the the
$(l=2, m=2)$ time component of the self-force we show the convergence of
our time-domain calculation to the frequency-domain result. We have
excellent convergence after a time of $200M$, which is approximately one
orbital period.

\begin{table}[h]
\centering
\begin{tabular}{c  c  c  c  c}
\hline\hline
{} & $R$ & Time-domain & Frequency-domain & error\\
\hline
$\partial_t\psi^R$ & $10M$ & $3.750211\times 10^{-5}$ & $3.750227\times 10^{-5}$ & 0.000431\% \\
$\partial_r\psi^R$ & $10M$ & $1.380612\times 10^{-5}$ & $1.378448\times 10^{-5}$ & 0.157\% \\
\hline
$\partial_t\psi^R$ & $12M$ & $1.747278\times 10^{-5}$ & $1.747254\times 10^{-5}$ & 0.00139\% \\
$\partial_r\psi^R$ & $12M$ & $5.715982\times 10^{-6}$ & $5.710205\times 10^{-6}$ & 0.101\% \\
\hline\hline
\end{tabular}
\caption{Summary of self-force results for $R=10M$ and $R=12M$. The error is
determined by a comparison with an accurate frequency-domain calculation
\cite{DetweilerMW2003}.}
\label{tab:only}
\end{table}

\section{Discussion}
\label{sec:summary}

In the specific context of a point charge orbiting a back hole, we have
introduced a very general approach suitable for the time-domain generation
of waveforms and also the calculation of the backreacting self-force.

Our initial tests are admittedly on the very restrictive case of circular
orbits of the Schwarzschild geometry, where we have taken advantage of the
spherical symmetry to decompose the source and field into spherical
harmonics. This has allowed us to compare our results to available
frequency-domain results of very high precision \cite{DetweilerMW2003,
Hikida2005}. In some manner, because of our use of a spherical-harmonic
decomposition, our analysis might be likened to using spectral methods.
But, the method of field regularization inherently does not require a
mode-decomposition and could be implemented with a full (3+1) numerical
code. For our test case we achieve an extremely accurate calculation for
the time component of the self-force $\partial_t\tilde\psi^R$, which is
equivalent to the rate of energy lost by radiation. Notably, in our (1+1)
implementation, the initial data settled down to provide this accurate
component of the self-force within only one orbit of the particle as shown
in Fig.~\ref{equilibrium}. This might be contrasted with a calculation of
$dE/dt$ made from a flux integral evaluated in the wave-zone, which, with
similarly unspecified initial data, requires evolution over a substantial
number of orbits.

For a circular orbit, the radial component of the self-force
$\partial_r\tilde\psi^R$ is conservative and generally more difficult to
calculate. We were able to match more accurate analyses
\cite{DetweilerMW2003, Hikida2005, HaasP2006, Haas2007} to about $0.1\%$.
With the spherical harmonic decomposition, our analysis went up to
$L = 39$. This relatively high number is due primarily to
the slow polynomial convergence resulting from the mode-decomposition of
the self-force. We expect this to be endemic in all self-force calculations
that rely on some kind of spectral decomposition, as it is the penalty
incurred when one represents objects of limited differentiability in terms of
smooth functions.

A technique similar to that described in Ref.~\cite{DetweilerMW2003} could
possibly mitigate this weakness. For our specific implementation, we could
choose to calculate and sum modes only up to, say $l=15$, and then take
advantage of the known asymptotic fall-off in $l$ shown in
Eq.~(\ref{regularization}). Using the computed modes, we determine the
coefficients in the expected fall-off for the self-force, and then using
these, analytically complete the sum to $l=\infty$. This results in a
slightly more accurate result for $\partial_r\tilde\psi^R$. A similar
procedure of ``fitting" to a known asymptotic fall-off might prove useful
if one chooses to implement field regularization using spectral methods.

We expect field regularization to be best implemented on a (3+1)
finite-difference code, with mesh refinement in the vicinity of the charge to
better resolve the limited differentiability of our analytically constructed
source function. Such a process will ameliorate the problem of slow
polynomial convergence ailing typical mode-sum prescriptions.

For the EMRI problem today, there is great interest in calculating the
rate of energy being radiated for a point mass orbiting a rotating black
hole and in using the result to modify the the orbit of the mass with some
version of an adiabatic approximation. For a general orbit, the energy
flux is not easy to determine. Current methods use the axial symmetry of
the Kerr geometry to separate out one dimension, and then deal with a
(2+1)D problem for the radiation from a point mass. The representation of
a point mass on a grid is typically problematical. Replacing a
$\delta$-function source by a narrow Gaussian \cite{Kahanna2003,
Kahanna2007} is reasonable but does not accurately reproduce
frequency-domain results. The recent distribution of a $\delta$-function
over a modest number of grid points by Sundararajan et al
\cite{Sundararajan2007} appears more robust. The strategy laid out here
provides a natural remedy to this issue. Instead of dealing with a wave
equation with a $\delta$-function source, we solve an \emph{equivalent}
problem with a regular and distributed source. The results displayed in
Figures \ref{TDFDcomp} and \ref{TDFDcompErr} clearly point to the
effectiveness and accuracy of our method.

Beyond the numerical modeling of $\delta$-function sources though, the method of
field regularization provides direct access to the self-force, which is essential
in a fully-consistent treatment of particle motion and wave generation. Current
methods under development are based upon energy and angular momentum flux
calculations that will certainly miss conservative self-force effects. These methods rely
upon flux integrals evaluated in the wave-zone and some orbit averaging or
post-processing to effect the change in orbital energy or angular momentum, which are
difficult to implement carefully \cite{PoundP2007a, PoundP2007b} and to justify
rigorously. The more direct approach of locally calculating the self-force to update
the particle orbit has been largely avoided because of the prohibitive computational
expense associated with mode-sum calculations of the self-force. In a
(3+1) finite-differencing implementation of field regularization, calculating
the self-force is no more expensive than performing a
numerical derivative and possibly an interpolation. As such, it represents
a step forward towards the goal of efficiently producing consistent numerical models
of particle motion and radiation in curved spacetime.

\begin{figure}[h]
  \centering
  \includegraphics[width=9cm,angle=0]{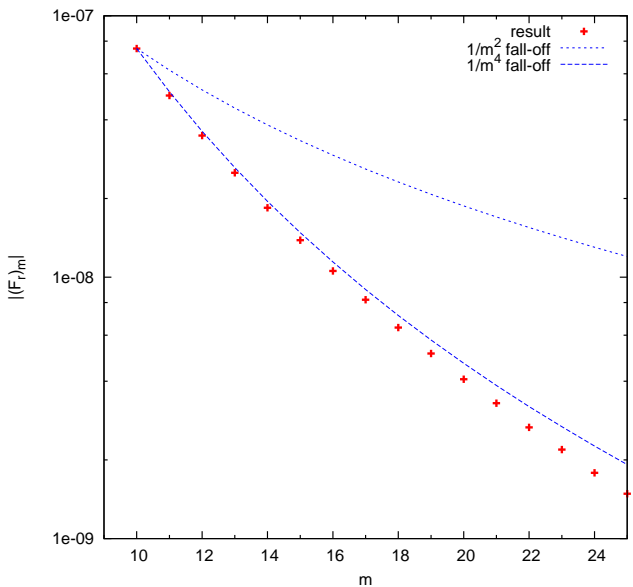}
\caption{$(\nabla_r\psi^R)_m$ versus $m$. Our results show an $m$-fall-off
closest to $1/m^4$.}
  \label{highM}
\end{figure}

The recent proposal of Barack, Golbourn and Sago \cite{BarackG2007,
BarackSago2007, BarackGS2007} is closest in spirit to our method of field
regularization. They model a point charge with a distributed effective
source derived instead from their ``puncture function'', which is quite similar to
our $\tilde\psi^S$.  They base their construction of the puncture function on the
`direct'+`tail' decomposition, rather than on the Green function decomposition
in \cite{DetweilerW2003} that naturally provides our regularizing singular field
$\tilde\psi^S$. Their current puncture function, however, appears to prevent them from
calculating a self-force. Moreover, in anticipation of a Kerr background application,
they envisage using a (2+1) code, necessitating a mode-sum over a mode index $m$, which
will again feature the characteristic polynomial convergence of this approach to
self-force calculation. This is demonstrated in Fig. \ref{highM}. Using our results,
we perform partial sums over $l$, i.e.
\begin{equation}
(\nabla_r\psi^R)_m \equiv \sum_{l>|m|}^{L} (\nabla_r\psi^R)_{lm},
\end{equation}
to get the resulting fall-off in $m$. We observe a fall-off close to $1/m^4$ in the
modes. Consequently, if we were to follow Barack et al's $m$-mode prescription, the
self-force would converge as $1/m^3$.

In principle, our method of field regularization appears to resolve two
important issues in the context of EMRI simulations: (a) numerically
representing $\delta$-function sources, and (b) calculating the
self-force. At this time our test of a scalar charge in a circular orbit
of the Schwarzschild geometry is a carefully controlled numerical
experiment and provides us with detailed information about the
relationship between the approximation for $\psi^S$ and the rate of
convergence of the self-force. However, our test is also extremely
elementary when compared to the actual case of a point mass emitting
gravitational waves from a generic orbit of the Kerr geometry, which is
most relevant for EMRIs. Future work will focus on exploring the
robustness of our technique against these more interesting cases.

\begin{acknowledgments}
We thank  Roland Haas, Eric Poisson, and Bernard Whiting for helpful
discussions pertaining to various aspects of this work. Development of the
ideas in this manuscript began while one of us (S.D.) was at the Aspen
Center for Physics during the 2005 summer workshop LISA Data: Analysis,
Sources, and Science; we gratefully acknowledge the Center and the
workshop organizers for their support and kind hospitality. These ideas
were further developed at the eighth, ninth and tenth Annual Capra
meetings at the Rutherford Appleton Laboratory Oxford (2005), University
of Wisconsin in Milwaukee (2006) and University of Alabama in Huntsville
(2007), respectively. And we are grateful to the organizers of these
valuable workshops. The authors also acknowledge the University of Florida
High-Performance Computing Center (URL: http://hpc.ufl.edu) and the
Institute for Fundamental Theory (URL: http://www.phys.ufl.edu/ift) for
providing computational resources and support that have contributed to the
research results reported in this paper. This work was supported in part
by the National Science Foundation, award No. PHY-0555484.

\end{acknowledgments}

\appendix

 \newcommand{\s}{{\text{s}}}
 \newcommand{\ro}{R}
\section{Approximate singular field in THZ-coordinates}
\label{sec:THZ} A special coordinate system developed by Thorne and Hartle
\cite{ThorneHartle85} and by Zhang \cite{Zhang86} is particularly useful
for self-force analyses. These THZ coordinates $(t,x,y,z)$ are defined in
a neighborhood of a geodesic of a vacuum spacetime and are locally
inertial, harmonic and Minkowskii-like, and centered on the geodesic with
$t$ measuring the proper time along the geodesic. In these special
coordinates our expression for the approximate singular field
$\tilde\psi_S = q/\sqrt{x^2+y^2+z^2}$ is quite simple. But, for the case
of a point charge in a circular orbit about a Schwarzschild black hole
this simplicity of $\psi^S$ belies the hidden complexity of the coordinate
transformation between the Schwarzschild coordinates $(t_s,r,\theta,\phi)$
and the THZ coordinates.

The full coordinate transformation, which we use for our analysis in the
main body of this paper may be found in Eqs.~(B1)-(B9) of
Ref.~\cite{DetweilerMW2003}. Below we give only an abbreviated form of
these equations to give a sense of how the coordinate transformation is
implemented. The formulae below give $\bar x$, $\bar y$, $\bar z$ and
$\bar t$ as smooth functions of the Schwarzschild $r$, $\theta$, $ \phi$
and $t_s$. We can then define a function $\bar\rho = \sqrt{\bar x^2+\bar
y^2+\bar z^2}$ which has the property that
\begin{equation}
  \nabla^a\nabla_a(1/\bar\rho) = -4\pi\delta(\vec x) + O(1/\bar\rho) .
\end{equation}
If we had used $q/\bar\rho$ as the approximation $\tilde\psi_s$ for the
singular field then the effective source for the regular field $\psi^R$ in
the vicinity of the point charge would be singular,
\begin{equation}
  \bar S_{\text{eff}} = -\nabla^a\nabla_b(q/\bar\rho) - 4\pi q \delta(\vec x) = O(1/\bar\rho),
\end{equation}
rather than $O(\rho)$, which is the case for the effective source which we
actually use as described in Eq.~(\ref{eq:SeffBehave}).

The coordinates which lead to $\bar\rho$ are now given for a circular
geodesic of the Schwarzschild geometry at Schwarzschild radius $\ro$: We
first define two useful functions
\begin{eqnarray}
   \tilde{x} &&= \frac{ [r \sin\theta \cos(\phi-\Omega t_\s)-\ro] }{(1-2M/\ro)^{1/2}}
  + \frac{M}{\ro^2(1-2M/\ro)^{1/2}}
\nonumber\\ && \times
    \left[\frac{(r-\ro)^2}{2(1-2M/\ro)} + \ro^2\sin^2\theta \sin^2(\phi-\Omega t_s) + \ro^2\cos^2\theta\right]
\nonumber \\&&
   \qquad\qquad {} +O(\rho^3)
\label{thzxbar}
\end{eqnarray}
 and
\begin{eqnarray}
 \tilde{y} &=&  r \sin\theta \sin(\phi-\Omega t_\s)
                              \left(\frac{\ro-2M}{\ro-3M}\right)^{1/2}\!\!\!\!\!\! + O(\rho^3) .
\label{thzybar}
\end{eqnarray}
The $O(\rho^3)$ terms indicate that these (and the formulae below) could
be modified by the addition of arbitrary $O(\rho^3)$ terms without
necessarily changing the usefulness of these coordinates.

In terms of these two functions, the THZ coordinates
 $(\bar t,\bar x,\bar y,\bar z)$ are
\begin{equation}
  \bar x = \tilde{x} \cos(\Omega^\dag t_\s) - \tilde{y} \sin(\Omega^\dag t_\s)
\label{thzx}
\end{equation}
and
\begin{equation}
  \bar y = \tilde{x} \sin(\Omega^\dag t_\s) + \tilde{y} \cos(\Omega^\dag t_\s)
\label{thzy}
\end{equation}
where $\Omega^\dag=\Omega\sqrt{1-3 M/\ro}$, along with
\begin{eqnarray}
 \bar z &=& r \cos(\theta)   + O(\rho^3)
\label{thzz}
\end{eqnarray}
and
\begin{eqnarray}
  \bar t &=& t_\s(1-3M/\ro)^{1/2}
  \nonumber\\ && \qquad{} - \frac{r\Omega \ro\sin\theta \sin(\phi-\Omega t_s)}{\ro-3M}  + O(\rho^3)
\label{thzt}
\end{eqnarray}

The set of functions $(\bar t,\tilde{x},\tilde{y},\bar z)$ forms a
non-inertial coordinate system that co-rotates with the particle in the
sense that the $\tilde{x}$ axis always lines up the center of the black
hole and the center of the particle, the $\tilde{y}$ axis is always
tangent to the spatially circular orbit, and the $\bar z$ axis is always
orthogonal to the orbital plane.

The THZ coordinates $(\bar t,\bar x,\bar y,\bar z)$ are locally inertial
and non-rotating in the vicinity of the charge, but these same coordinates
appear to be rotating when viewed far from the charge as a consequence of
Thomas precession as revealed in the $\Omega^\dag t_\s$ dependence in
Eqs.~(\ref{thzx}) and (\ref{thzy}) above.

The coordinates $(\bar t,\bar x,\bar y,\bar z)$ given above are said to be
\textit{second order} THZ coordinates and differ from the actual
\textit{fourth order} ones used in the main body of this paper by the
replacement of the $O(\rho^3)$ terms appearing above by specific terms
which scale as $\rho^3$ and $\rho^4$ \cite{DetweilerMW2003} and leave the
undetermined parts of the THZ coordinates being $O(\rho^5)$.\vspace{1em}

\section{Convergence factor}
\label{sec:convergence}

An $n$th-order evolution code is one for which $\psi = \psi_N(h) +(\epsilon^{(n)}(\xi))h^n$ ,
where $\psi_N(h)$ is the numerical solution at resolution $h$, and $\epsilon^{(n)}(\xi)$ is some unknown
error function or order $\approx 1$. Consider three resolutions $h$, $2h$, $4h$. This then leads to
\begin{align}
\psi &= \psi_N(h) +(\epsilon^{(n)}(\xi))h^n \\
\psi &= \psi_N(2h) +(\epsilon^{(n)}(\xi))(2h)^n\\
\psi &= \psi_N(4h) +(\epsilon^{(n)}(\xi))(4h)^n
\end{align}
Thus,
\begin{align}
\frac{|\psi_N(4h)-\psi_N(2h)|}{|\psi_N(2h)-\psi_N(h)|}= |\epsilon^{(n)}(\xi)|2^n,
\end{align} and so
\begin{align}
n = \log &\left|\frac{\psi_N(4h)-\psi_N(2h)}{\psi_N(2h)-\psi_N(h)}\right|/\log(2) \nonumber \\ &+ \log|\epsilon^{(n)}(\xi)|/\log(2).
\end{align}


\begin{thebibliography}{33}
\expandafter\ifx\csname natexlab\endcsname\relax\def\natexlab#1{#1}\fi
\expandafter\ifx\csname bibnamefont\endcsname\relax
  \def\bibnamefont#1{#1}\fi
\expandafter\ifx\csname bibfnamefont\endcsname\relax
  \def\bibfnamefont#1{#1}\fi
\expandafter\ifx\csname citenamefont\endcsname\relax
  \def\citenamefont#1{#1}\fi
\expandafter\ifx\csname url\endcsname\relax
  \def\url#1{\texttt{#1}}\fi
\expandafter\ifx\csname urlprefix\endcsname\relax\def\urlprefix{URL }\fi
\providecommand{\bibinfo}[2]{#2} \providecommand{\eprint}[2][]{\url{#2}}

\bibitem[{\citenamefont{Regge and Wheeler}(1957)}]{ReggeWheeler}
    \bibinfo{author}{\bibfnamefont{T.}~\bibnamefont{Regge}}
    \bibnamefont{and}
  \bibinfo{author}{\bibfnamefont{J.~A.} \bibnamefont{Wheeler}},
  \bibinfo{journal}{Phys. Rev.} \textbf{\bibinfo{volume}{108}},
  \bibinfo{pages}{1063} (\bibinfo{year}{1957}).

\bibitem[{\citenamefont{Zerilli}(1970)}]{Zerilli}
    \bibinfo{author}{\bibfnamefont{F.~J.} \bibnamefont{Zerilli}},
  \bibinfo{journal}{Phys. Rev. D} \textbf{\bibinfo{volume}{2}},
  \bibinfo{pages}{2141} (\bibinfo{year}{1970}).

\bibitem[{\citenamefont{Teukolsky}(1973)}]{Teukolsky73}
    \bibinfo{author}{\bibfnamefont{S.~A.} \bibnamefont{Teukolsky}},
  \bibinfo{journal}{Astrophys. J.} \textbf{\bibinfo{volume}{185}},
  \bibinfo{pages}{635} (\bibinfo{year}{1973}).

\bibitem[{\citenamefont{{L. Barack and A. Ori}}(2000)}]{BarackO2000}
    \bibinfo{author}{\bibnamefont{{L. Barack and A. Ori}}},
    \bibinfo{journal}{Phys.
  Rev. D} \textbf{\bibinfo{volume}{61}}, \bibinfo{pages}{061502}
  (\bibinfo{year}{2000}), \eprint{http://arxiv.org/abs/gr-qc/9912010}.

\bibitem[{\citenamefont{Barack et~al.}(2002)\citenamefont{Barack, Mino,
    Nakano,
  Ori, and Sasaki}}]{BMNOS02}
\bibinfo{author}{\bibfnamefont{L.}~\bibnamefont{Barack}},
  \bibinfo{author}{\bibfnamefont{Y.}~\bibnamefont{Mino}},
  \bibinfo{author}{\bibfnamefont{H.}~\bibnamefont{Nakano}},
  \bibinfo{author}{\bibfnamefont{A.}~\bibnamefont{Ori}}, \bibnamefont{and}
  \bibinfo{author}{\bibfnamefont{M.}~\bibnamefont{Sasaki}},
  \bibinfo{journal}{Phys. Rev. Lett.} \textbf{\bibinfo{volume}{88}},
  \bibinfo{pages}{091101} (\bibinfo{year}{2002}).

\bibitem[{\citenamefont{Mino et~al.}(2002)\citenamefont{Mino, Nakano, and
  Sasaki}}]{MinoNS2002}
\bibinfo{author}{\bibfnamefont{Y.}~\bibnamefont{Mino}},
  \bibinfo{author}{\bibfnamefont{H.}~\bibnamefont{Nakano}}, \bibnamefont{and}
  \bibinfo{author}{\bibfnamefont{M.}~\bibnamefont{Sasaki}},
  \bibinfo{journal}{Prog. Theor. Phys.} \textbf{\bibinfo{volume}{108}},
  \bibinfo{pages}{1039} (\bibinfo{year}{2002}),
  \eprint{http://arxiv.org/abs/gr-qc/0111074}.

\bibitem[{\citenamefont{Lousto}(2000)}]{Lousto00}
    \bibinfo{author}{\bibfnamefont{C.~O.} \bibnamefont{Lousto}},
  \bibinfo{journal}{Phys. Rev. Lett.} \textbf{\bibinfo{volume}{84}},
  \bibinfo{pages}{5251} (\bibinfo{year}{2000}).

\bibitem[{\citenamefont{Burko}(2000)}]{Burko00}
    \bibinfo{author}{\bibfnamefont{L.~M.} \bibnamefont{Burko}},
  \bibinfo{journal}{Phys. Rev. Lett.} \textbf{\bibinfo{volume}{84}},
  \bibinfo{pages}{4529} (\bibinfo{year}{2000}).

\bibitem[{\citenamefont{{S. Detweiler, E. Messaritaki, and B.F.
  Whiting}}(2003)}]{DetweilerMW2003}
\bibinfo{author}{\bibnamefont{{S. Detweiler, E. Messaritaki, and B.F.
  Whiting}}}, \bibinfo{journal}{Phys. Rev. D} \textbf{\bibinfo{volume}{67}},
  \bibinfo{pages}{104016} (\bibinfo{year}{2003}),
  \eprint{http://arxiv.org/abs/gr-qc/0205079}.

\bibitem[{\citenamefont{Diaz-Rivera
    et~al.}(2004)\citenamefont{Diaz-Rivera,
  Messaritaki, Whiting, and Detweiler}}]{RiveraMWD2004}
\bibinfo{author}{\bibfnamefont{L.~M.} \bibnamefont{Diaz-Rivera}},
  \bibinfo{author}{\bibfnamefont{E.}~\bibnamefont{Messaritaki}},
  \bibinfo{author}{\bibfnamefont{B.}~\bibnamefont{Whiting}}, \bibnamefont{and}
  \bibinfo{author}{\bibfnamefont{S.}~\bibnamefont{Detweiler}},
  \bibinfo{journal}{Phys. Rev. D} \textbf{\bibinfo{volume}{70}},
  \bibinfo{pages}{124018} (\bibinfo{year}{2004}),
  \eprint{http://arxiv.org/abs/gr-qc/0410011}.

\bibitem[{\citenamefont{Hikida et~al.}(2004)\citenamefont{Hikida, Jhingan,
  Nakano, Sago, Sasaki, and Tanaka}}]{Hikida2004}
\bibinfo{author}{\bibfnamefont{W.}~\bibnamefont{Hikida}},
  \bibinfo{author}{\bibfnamefont{S.}~\bibnamefont{Jhingan}},
  \bibinfo{author}{\bibfnamefont{H.}~\bibnamefont{Nakano}},
  \bibinfo{author}{\bibfnamefont{N.}~\bibnamefont{Sago}},
  \bibinfo{author}{\bibfnamefont{M.}~\bibnamefont{Sasaki}}, \bibnamefont{and}
  \bibinfo{author}{\bibfnamefont{T.}~\bibnamefont{Tanaka}},
  \bibinfo{journal}{Prog. Theor. Phys.} \textbf{\bibinfo{volume}{111}},
  \bibinfo{pages}{821} (\bibinfo{year}{2004}),
  \eprint{http://arxiv.org/abs/gr-qc/0308068}.

\bibitem[{\citenamefont{Hikida et~al.}(2005)\citenamefont{Hikida, Jhingan,
  Nakano, Sago, Sasaki, and Tanaka}}]{Hikida2005}
\bibinfo{author}{\bibfnamefont{W.}~\bibnamefont{Hikida}},
  \bibinfo{author}{\bibfnamefont{S.}~\bibnamefont{Jhingan}},
  \bibinfo{author}{\bibfnamefont{H.}~\bibnamefont{Nakano}},
  \bibinfo{author}{\bibfnamefont{N.}~\bibnamefont{Sago}},
  \bibinfo{author}{\bibfnamefont{M.}~\bibnamefont{Sasaki}}, \bibnamefont{and}
  \bibinfo{author}{\bibfnamefont{T.}~\bibnamefont{Tanaka}},
  \bibinfo{journal}{Prog. Theor. Phys.} \textbf{\bibinfo{volume}{113}},
  \bibinfo{pages}{283} (\bibinfo{year}{2005}),
  \eprint{http://arxiv.org/abs/gr-qc/0410115}.

\bibitem[{\citenamefont{Haas and Poisson}(2006)}]{HaasP2006}
    \bibinfo{author}{\bibfnamefont{R.}~\bibnamefont{Haas}}
    \bibnamefont{and}
  \bibinfo{author}{\bibfnamefont{E.}~\bibnamefont{Poisson}},
  \bibinfo{journal}{Physical Review D} \textbf{\bibinfo{volume}{74}},
  \bibinfo{pages}{044009} (\bibinfo{year}{2006}),
  \eprint{http://arXiv.org/abs/gr-qc/0605077}.

\bibitem[{\citenamefont{{R. Haas}}(2007)}]{Haas2007}
    \bibinfo{author}{\bibnamefont{{R. Haas}}}, \bibinfo{journal}{Phys.
    Rev. D}
  \textbf{\bibinfo{volume}{75}}, \bibinfo{pages}{124011}
  (\bibinfo{year}{2007}), \\ \eprint{http://arxiv.org/abs/gr-qc/0704.0797}.

\bibitem[{\citenamefont{Barack and Sago}(2007)}]{BarackSago2007}
    \bibinfo{author}{\bibfnamefont{L.}~\bibnamefont{Barack}}
    \bibnamefont{and}
  \bibinfo{author}{\bibfnamefont{N.}~\bibnamefont{Sago}},
  \bibinfo{journal}{Phys. Rev. D} \textbf{\bibinfo{volume}{75}},
  \bibinfo{pages}{064021} (\bibinfo{year}{2007}),
  \eprint{http://arxiv.org/abs/gr-qc/0701069}.

\bibitem[{\citenamefont{{B.S. DeWitt and R.W.
    Brehme}}(1960)}]{DeWittB1960} \bibinfo{author}{\bibnamefont{{B.S.
    DeWitt and R.W. Brehme}}},
  \bibinfo{journal}{Ann. Phys.} \textbf{\bibinfo{volume}{9}},
  \bibinfo{pages}{220} (\bibinfo{year}{1960}).

\bibitem[{\citenamefont{{Y. Mino, M. Sasaki, T.
    Tanaka}}(1997)}]{MinoST1997} \bibinfo{author}{\bibnamefont{{Y. Mino,
    M. Sasaki, T. Tanaka}}},
  \bibinfo{journal}{Phys. Rev. D} \textbf{\bibinfo{volume}{55}},
  \bibinfo{pages}{3457} (\bibinfo{year}{1997}),
  \eprint{http://arxiv.org/abs/gr-qc/9606018}.

\bibitem[{\citenamefont{{T.C. Quinn and R.M. Wald}}(1997)}]{QuinnW1997}
    \bibinfo{author}{\bibnamefont{{T.C. Quinn and R.M. Wald}}},
  \bibinfo{journal}{Phys. Rev. D} \textbf{\bibinfo{volume}{56}},
  \bibinfo{pages}{3381} (\bibinfo{year}{1997}),
  \eprint{http://arxiv.org/abs/gr-qc/9610053}.

\bibitem[{\citenamefont{{S. Detweiler and B.F.
  Whiting}}(2003)}]{DetweilerW2003}
\bibinfo{author}{\bibnamefont{{S. Detweiler and B.F. Whiting}}},
  \bibinfo{journal}{Phys. Rev. D} \textbf{\bibinfo{volume}{67}},
  \bibinfo{pages}{024025} (\bibinfo{year}{2003}),
  \eprint{http://arxiv.org/abs/gr-qc/0202086}.

\bibitem[{\citenamefont{Thorne and Hartle}(1985)}]{ThorneHartle85}
    \bibinfo{author}{\bibfnamefont{K.~S.} \bibnamefont{Thorne}}
    \bibnamefont{and}
  \bibinfo{author}{\bibfnamefont{J.~B.} \bibnamefont{Hartle}},
  \bibinfo{journal}{Phys. Rev. D} \textbf{\bibinfo{volume}{31}},
  \bibinfo{pages}{1815} (\bibinfo{year}{1985}).

\bibitem[{\citenamefont{Zhang}(1986)}]{Zhang86}
    \bibinfo{author}{\bibfnamefont{X.-H.} \bibnamefont{Zhang}},
  \bibinfo{journal}{Phys. Rev. D} \textbf{\bibinfo{volume}{34}},
  \bibinfo{pages}{991} (\bibinfo{year}{1986}).

\bibitem[{\citenamefont{Detweiler}(2001)}]{det01}
    \bibinfo{author}{\bibfnamefont{S.}~\bibnamefont{Detweiler}},
  \bibinfo{journal}{Phys. Rev. Lett.} \textbf{\bibinfo{volume}{86}},
  \bibinfo{pages}{1931} (\bibinfo{year}{2001}),
  \eprint{http://arXiv.org/abs/grqc/0011039}.

\bibitem[{\citenamefont{Detweiler}(2005)}]{Detweiler05}
    \bibinfo{author}{\bibfnamefont{S.}~\bibnamefont{Detweiler}},
  \bibinfo{journal}{Class. Quantum Grav.} \textbf{\bibinfo{volume}{22}},
  \bibinfo{pages}{S681} (\bibinfo{year}{2005}),
  \eprint{http://arxiv.org/abs/gr-qc/0501004}.

\bibitem[{\citenamefont{Press et~al.}(1992)\citenamefont{Press, Teukolsky,
  Vetterling, and Flannery}}]{NumericalRecipes}
\bibinfo{author}{\bibfnamefont{W.~H.} \bibnamefont{Press}},
  \bibinfo{author}{\bibfnamefont{S.~A.} \bibnamefont{Teukolsky}},
  \bibinfo{author}{\bibfnamefont{W.~T.} \bibnamefont{Vetterling}},
  \bibnamefont{and} \bibinfo{author}{\bibfnamefont{B.~P.}
  \bibnamefont{Flannery}}, \emph{\bibinfo{title}{Numerical Recipes in {C}: The
  Art of Scientific Computing}} (\bibinfo{publisher}{Cambridge University
  Press}, \bibinfo{address}{Cambridge}, \bibinfo{year}{1992}),
  \bibinfo{edition}{2nd} ed.

\bibitem[{\citenamefont{{C.O. Lousto and R.H. Price}}(1997)}]{LoustoP1997}
    \bibinfo{author}{\bibnamefont{{C.O. Lousto and R.H. Price}}},
  \bibinfo{journal}{Phys. Rev. D} \textbf{\bibinfo{volume}{56}},
  \bibinfo{pages}{6439} (\bibinfo{year}{1997}),
  \eprint{http://arxiv.org/abs/gr-qc/9705071}.

\bibitem[{\citenamefont{{C.O. Lousto}}(2005)}]{Lousto2005}
    \bibinfo{author}{\bibnamefont{{C.O. Lousto}}},
    \bibinfo{journal}{Class. Quantum
  Grav.} \textbf{\bibinfo{volume}{22}}, \bibinfo{pages}{S543}
  (\bibinfo{year}{2005}), \eprint{http://arxiv.org/abs/gr-qc/0503001}.

\bibitem[{\citenamefont{Lopez-Aleman
    et~al.}(2003)\citenamefont{Lopez-Aleman,
  Khanna, and Pullin}}]{Kahanna2003}
\bibinfo{author}{\bibfnamefont{R.}~\bibnamefont{Lopez-Aleman}},
  \bibinfo{author}{\bibfnamefont{G.}~\bibnamefont{Khanna}}, \bibnamefont{and}
  \bibinfo{author}{\bibfnamefont{J.}~\bibnamefont{Pullin}},
  \bibinfo{journal}{Classical and Quantum Gravity}
  \textbf{\bibinfo{volume}{20}}, \bibinfo{pages}{3259} (\bibinfo{year}{2003}),
  \eprint{http://arxiv.org/abs/gr-qc/0303054}.

\bibitem[{\citenamefont{Burko and Khanna}(2007)}]{Kahanna2007}
    \bibinfo{author}{\bibfnamefont{L.~M.} \bibnamefont{Burko}}
    \bibnamefont{and}
  \bibinfo{author}{\bibfnamefont{G.}~\bibnamefont{Khanna}},
  \bibinfo{journal}{Europhysics Letters} \textbf{\bibinfo{volume}{78}},
  \bibinfo{pages}{60005} (\bibinfo{year}{2007}),
  \eprint{http://arXiv.org/abs/gr-qc/0609002}.

\bibitem[{\citenamefont{Sundararajan
    et~al.}(2007)\citenamefont{Sundararajan,
  Khanna, and Hughes}}]{Sundararajan2007}
\bibinfo{author}{\bibfnamefont{P.~A.} \bibnamefont{Sundararajan}},
  \bibinfo{author}{\bibfnamefont{G.}~\bibnamefont{Khanna}}, \bibnamefont{and}
  \bibinfo{author}{\bibfnamefont{S.~A.} \bibnamefont{Hughes}},
  \bibinfo{journal}{Phys. Rev. D} \textbf{\bibinfo{volume}{76}},
  \bibinfo{pages}{104005} (\bibinfo{year}{2007}).

\bibitem[{\citenamefont{Pound and
    Poisson}(2007{\natexlab{a}})}]{PoundP2007a}
    \bibinfo{author}{\bibfnamefont{A.}~\bibnamefont{Pound}}
    \bibnamefont{and}
  \bibinfo{author}{\bibfnamefont{E.}~\bibnamefont{Poisson}}
  (\bibinfo{year}{2007}{\natexlab{a}}), \\ \eprint{http://arXiv.org/abs/0708.3037
  [gr-qc]}.

\bibitem[{\citenamefont{Pound and
    Poisson}(2007{\natexlab{b}})}]{PoundP2007b}
    \bibinfo{author}{\bibfnamefont{A.}~\bibnamefont{Pound}}
    \bibnamefont{and}
  \bibinfo{author}{\bibfnamefont{E.}~\bibnamefont{Poisson}}
  (\bibinfo{year}{2007}{\natexlab{b}}), \\ \eprint{http://arXiv.org/abs/0708.3033
  [gr-qc]}.

\bibitem[{\citenamefont{{L. Barack and D. Golbourn}}(2007)}]{BarackG2007}
    \bibinfo{author}{\bibnamefont{{L. Barack and D. Golbourn}}},
  \bibinfo{journal}{Phys. Rev. D} \textbf{\bibinfo{volume}{76}},
  \bibinfo{pages}{044020} (\bibinfo{year}{2007}),
  \eprint{http://arxiv.org/abs/gr-qc/0705.3620}.

\bibitem[{\citenamefont{Barack et~al.}(2007)\citenamefont{Barack,
    Golbourn, and
  Sago}}]{BarackGS2007}
\bibinfo{author}{\bibfnamefont{L.}~\bibnamefont{Barack}},
  \bibinfo{author}{\bibfnamefont{D.~A.} \bibnamefont{Golbourn}},
  \bibnamefont{and} \bibinfo{author}{\bibfnamefont{N.}~\bibnamefont{Sago}}
  (\bibinfo{year}{2007}), \\ \eprint{http://arXiv.org/abs/0709.4588 [gr-qc]}.

\end{thebibliography}

\end{document}